# A Four-Level Ontological Framework for Quantum Field Theory: From Quantum Vacuum to Phenomenal Reality

Ali Reza Mirzaee(Mirza), Tehran University[1]

## Abstract

This paper proposes a four-level ontological framework for understanding the structure of reality as implied by Quantum Field Theory (QFT). The four levels are: the Quantum Vacuum(Level 0), the Virtual (Level 1), the Quantum or Actualizable Level (Level 2 ), and the Phenomenal or Observable Reality (Level 3). Each level corresponds to a distinct mode of existence, ranging from the fundamental vacuum state to the emergent phenomenal domain. Within this framework, quantum fields are treated as the continuous ontological background giving rise to both virtual and actualizable entities. Virtual particles—those satisfying the mathematical structure of the theory but not directly observable—occupy Level 1, while quantum entities that satisfy Einstein's mass-energy relation belong to Level 2 .The transition from the virtual to the quantum level describes the process by which potentialities become actualities, offering a coherent view that integrates physical and philosophical interpretations of QFT. This model aims to dissolve the traditional dichotomy of "real" and "unreal" by situating both within a stratified and interconnected ontology.

## Introduction

One of the foundational challenges in both Quantum Field Theory (QFT) and the philosophy of physics lies in the traditional binary classification of entities as either real or unreal. The classical view posits that material entities are real, while mental constructs or theoretical entities are unreal. However, quantum phenomena such as virtual particles challenge this simplistic dichotomy. These unobservable entities, by influencing measurable phenomena, raise profound questions about the nature and modes of existence. To address this, we propose a four-level ontological hierarchy that spans from the quantum vacuum and the virtual level to the quantum level and phenomenal reality, providing a coherent continuum from potentiality to actualization. This framework not only clarifies the interpretation of quantum theory but also offers a unified perspective linking the unobservable with the observable.

In QFT, virtual particles play a crucial role in mediating forces and in correcting the measurable properties of real particles. They arise within the mathematical structure of Feynman diagrams and, despite being unobservable, produce real, measurable effects such as the Lamb shift and the Casimir effect. Philosophically, this raises deep ontological questions: if unobservable entities affect measurable phenomena, can they truly be dismissed as unreal? Conversely, if they are real, what mode of existence do they possess? The traditional dichotomy of real versus unreal is thus insufficient for interpreting the ontological implications of quantum processes. To overcome these limitations, we propose a four-level ontological hierarchy.

---

[1] Mirzaee.alireza@ut.ac.ir

Level 0 (Quantum Vacuum): the fundamental field of potentiality, an undifferentiated ground underlying all phenomena.

Level 1 (Virtual Level): encompasses virtual particles arising from quantum fluctuations and mathematical structures, which do not satisfy the Einstein mass-energy relation.

Level 2 (Quantum Level): entities that are actualizable, satisfying the mass-energy relation and potentially observable under appropriate conditions.

Level 3 (Phenomenal Reality): the emergent macroscopic domain of observation and measurement.

This hierarchical framework provides a continuous ontological gradient from pure potentiality to phenomenal actuality, offering a unified account of both physical and philosophical aspects of quantum theory. Historically, attempts to resolve the question of reality in quantum theory have oscillated between strict instrumentalism and ontological realism. From Bohr's Copenhagen interpretation, which restricted reality to measurable phenomena, to Bohm's implicate order, which posited a deeper quantum reality, every approach struggled to reconcile the unobservable with the observable. The proposed four-level model follows this historical trajectory but reinterprets it as an ontological continuum rather than an epistemic divide. For instance, Knox (2023) explores spacetime emergence in QFT, emphasizing that the structure of fields implies layered ontological dependence. Fraser (2020) and Wallace (2021) examine the metaphysical commitments of QFT, arguing that entities like virtual particles resist clear classification within classical categories of existence. Ladyman and Lorenzetti (2024) suggest a structural realist view in which reality is constituted by relations rather than objects. French (2012) and Arntzenius (2018) investigate ontological pluralism in physics, showing that different levels of physical description correspond to different kinds of existence. Finally, Barad (2019) and Esfeld and Valentini (2022) emphasize the relational ontology of quantum phenomena, aligning with the idea that being itself is dynamic and context-dependent.

These contemporary analyses reinforce the necessity of a multi-level ontological approach. By integrating their insights, our model of four ontological levels—Vacuum, Virtual, Quantum, and Phenomenal—provides a framework in which both the mathematical structure of QFT and its experimental implications coexist coherently, without reducing one to the other.

1. Theorical Framework: The Four- Level Reality

The traditional division between real and unreal entities has long dominated both physics and philosophy. In classical ontology, only tangible, measurable objects are considered real, while theoretical or mental constructs are relegated to the status of non-being. However, this binary realism collapses under the weight of quantum theory, where entities such as virtual particles or quantum potentials appear to exert physical influence without being directly observable. Quantum mechanics and QFT demonstrate that something can be causally effective without being empirically accessible, which undermines the notion that only the measurable is real.

To move beyond this dichotomy, it becomes necessary to treat reality not as a single homogeneous domain but as a stratified structure of existence. The hierarchy proposed here begins with the Quantum Vacuum (Level 0), which grounds all possible states, and rises through successive levels of ontological actualization until reaching the Phenomenal Reality (Level 3) — the world of empirical phenomena. Each level is not separate from the others but emerges through a process of ontological unfolding, linking potentiality to actuality in a continuous manner.

At Level 0, the Quantum Vacuum represents the foundational state of pure potentiality. Contrary to classical intuition, this vacuum is not empty but filled with fluctuating quantum fields whose excitations are the seeds of all other levels. It provides the ontological substrate upon which all higher levels depend. The vacuum embodies a form of proto-being, containing all possible field configurations and virtual processes in latent form. Philosophically, this level corresponds to an undifferentiated ground of being—an ontic background that is not observable but ontologically necessary.

Emerging from the vacuum, Level 1 (Virtual) consists of entities that participate in the dynamics of quantum fields but do not satisfy the Einstein mass–energy relation. Virtual particles exist within the mathematical structure of QFT and mediate interactions between real, observable particles. They are not directly detectable yet are indispensable for explaining measurable effects such as the Lamb shift, vacuum polarization, and the Casimir effect. Their existence therefore cannot be dismissed as mere fiction; rather, it must be understood as a form of ontological semi-reality—entities that possess causal efficacy within theoretical structure without empirical manifestation.

At Level 2, quantum entities satisfy the Einstein relation , meaning they possess definable mass and energy. However, they exist only as potentially observable states until interaction or measurement brings them into phenomenality. This level bridges the virtual and the phenomenal: it is where probability amplitudes and wave functions represent tendencies toward actuality. Philosophically, Level 2 corresponds to actualizability — existence poised on the threshold of manifestation.

Level 3, the Phenomenal Level, is the domain of empirical observation — the world as experienced and measured. Here, the quantum potentialities of Level 2 actualize into determinate phenomena, giving rise to the macroscopic order of classical physics. Phenomenal Reality is not an independent layer detached from the underlying levels; it is an emergent expression of them, shaped by measurement, decoherence, and relational processes. Thus, what we call "reality" in the everyday sense is the final stage in a hierarchy that begins with the vacuum itself.

One of the key virtues of this four-level ontology is that it preserves causal continuity without collapsing the distinctions between levels. The vacuum (Level 0) provides the potential foundation; the virtual domain (Level 1) encodes intermediate fluctuations; the quantum level (Level 2) embodies actualizable states; and the phenomenal level (Level 3) manifests the observable consequences. Causality thus becomes a process of ontological propagation — from potential to virtual, from virtual to quantum, and from quantum to phenomenal. This framework replaces the rigid boundary between "real" and "unreal" with a fluid continuum of being.

By adopting this stratified ontology, Quantum Field Theory can be interpreted not merely as a computational framework but as an ontological map of existence. It offers a unified understanding of why unobservable entities such as virtual particles are both mathematically necessary and ontologically meaningful. Philosophically, it aligns with process metaphysics, structural realism, and relational ontologies that view being as dynamic and emergent. The four-level structure; Quantum Vacuum, Virtual, Quantum, and Phenomenal, thus provides a coherent language for bridging the conceptual gap between physics and metaphysics, dissolving the false dichotomy between the "real" and the "merely mathematical."

## 2. Particle Interpretation of Quantum Field Theory

One of the central difficulties in explaining physical and philosophical phenomena within the traditional dualistic framework of "real" versus "unreal" entities emerges when we examine the causal conditions underlying transitions between successive levels of reality. At the quantum level, this difficulty becomes evident in the necessity—imposed by the axioms of quantum mechanics—of accepting the principle of superposition. According to this principle, the wave function of a particle represents the linear superposition of all its possible states in a given Hilbert space. In a two-state spin system, for example, the electron's wave function prior to measurement encompasses both spin-up and spin-down components. Upon measurement, however, the wave function collapses into one of these states, a process often described as wave-function collapse.

The acceptance of superposition introduced a profound shift in the philosophical interpretation of probability. In classical contexts, probability was regarded as a subjective measure, reflecting incomplete knowledge of an otherwise determinate system—such as the uncertainty in predicting the outcome of a coin toss due to limited information about initial conditions. Quantum mechanics, however, requires an objective interpretation of probability: even with complete knowledge of all boundary conditions, the outcome remains intrinsically indeterminate. This view, famously resisted by Einstein's assertion that "God does not play dice," implies that indeterminacy is not epistemic but ontological—it arises from the very structure of nature rather than from our ignorance.

Consequently, quantum theory compels us to accept that a particle, prior to measurement, does not occupy a single definite state but rather exists as a distributed entity across all possible configurations of its wave function. Measurement, therefore, represents a non-causal and discontinuous transition from a manifold of possibilities to a single actualization. Such a view undermines both causality and determinism, raising the deeper question: how do macroscopic, phenomenally stable objects—with well-defined positions and properties—emerge from quantum constituents whose very nature is to exist without definite localization or classical identity?

Although the connection between quantum particles—lacking individuality, determinacy, and distinct identity—and phenomenal objects—characterized by spatial-temporal localization, distinctness, and predictability—can, to some extent, be explained through statistical mechanics applied to large ensembles of quantum particles, such explanations remain valid only at the phenomenal level. At the quantum level, fundamental issues such as indeterminacy, the breakdown of causality, and the problem of identity persist, revealing that the transition from quantum indeterminacy to phenomenal determinacy cannot be fully accounted for by mere statistical averaging. This suggests the necessity of a deeper conceptual framework to understand the ontological and causal relation between these two levels of reality.We suggest that these issues can be resolved by positing deeper ontological layers of reality, where the concept of "reality" itself attains a distinct meaning at each level.

To initiate our examination of the relationship between observable quantum particles—which are on-shell and satisfy Einstein's mass–energy relation—and virtual particles, which are off-shell and do not, we begin with the interaction of electron–positron annihilation and photon pair creation. The ontological status of virtual particles remains a matter of ongoing debate among both physicists and philosophers of physics. Many regard them as non-real entities, serving merely as mathematical constructs within the formalism of the theory (see Redhead, 1987; Weingard, 1987). The ontological status of virtual particles remains debated. While they appear in Feynman diagrams and affect interaction amplitudes, many studies treat them as mathematical tools rather than physically real entities (Jaeger, 2019; Anselmi, 2020; Zichert & Wüthrich, 2024). They function as calculational intermediaries, enabling precise predictions in quantum field theory, without implying direct observability.

### 3. Virtual Particles Reality

Figure 1 depicts the Feynman diagram for electron–positron annihilation and the subsequent creation of a photon pair. Prior to the ultimate collision, the quanta of the spinor fields—the electron and positron—are real (on-shell), meaning they are both experimentally observable and comply with Einstein's mass–energy relation. The intermediate particle connecting the two spinor fields is a virtual photon, representing a quantum of the electromagnetic gauge field. This photon is off-shell, as it neither satisfies the mass–energy relation nor can it be directly detected. At the precise moment of interaction, the configuration of the Feynman diagram transforms, as illustrated in Figure 2 (see Mandel, 1987, p. 112). Before the interaction, the mediating particle is a virtual photon, classified as off-shell. At the precise moment of collision, its status suddenly changes, and the mediator becomes a real electron; simultaneously, the electron, previously an on-shell spinor field quantum, transforms into a virtual, off-shell particle. The key question is how to interpret this mutual conversion. If, as some physicists and philosophers assume, virtual particles are merely mathematical tools with no independent existence, then how can one explain that a real particle becomes a "mathematical entity" and vice versa?

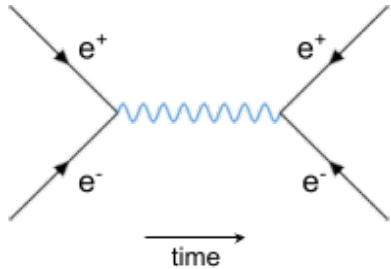

Figure 1, Feynman diagram for electron−positron annihilation and the subsequent creation of a photon pair. Prior to the ultimate collision. ***Propagator is a virtual photon***. ( Copied from the internet)

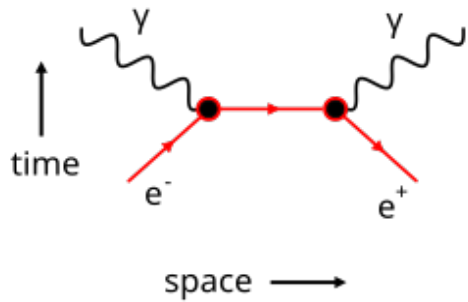

Figure 2, Feynman diagram for electron−positron annihilation and the subsequent creation of a photon pair. At the precise moment of interaction. ***Propagator is virtual electron***. ( Copied from the internet)

It is more consistent to assume that virtual particles are not merely mathematical tools, but rather bear a relation to a distinct form of reality. This form of reality differs from the on-shell quantum particles and also from phenomenal (observed) reality, yet it nonetheless constitutes a genuine mode of existence. In this framework, the conversion of a real particle into a virtual one, and vice versa, is not a simple transition between "real" and "non-real"; rather, it is a passage across different levels of reality, each governed by its own definitions and constraints.

At the phenomenal level, reality is defined through individuality and spatial localization as well as temporal continuity. In contrast, quantum particles lack definite positions in space and do not exhibit meaningful continuity in time; hence, the definition of reality for these entities no longer relies on localization or persistence, but rather on their potential observability in experiments. Virtual particles possess neither spatial-temporal definiteness nor direct experimental detectability, as they do not satisfy Einstein's mass-energy relation. Consequently, the definitions of reality from the preceding levels do not apply to this domain. However, in interactions such as electron-positron annihilation and photon pair creation, virtual particles demonstrate properties that can serve as a basis for defining reality at this level: the reality of virtual particles can be understood as their capacity to transition to the quantum, observable level. In other words, the reality of virtual particles is their potential observability. In the electron−positron annihilation and photon-pair creation, electrons and positrons accelerate toward each other under the electromagnetic field. At the Virtual Level (Level 1), particles may temporarily occupy superluminal or tachyonic-like states—features confined to the virtual domain. Virtual photons approach the speed of light and lose virtual mass, becoming real, on-shell photons, while the electron and positron transition to off-shell, virtual states. This illustrates the ontological passage between Virtual (Level 1) and Quantum (Level 2) domains, The superluminal behavior occurs only at the Virtual Level, while consistency with physical laws is preserved at the Phenomenal Level (Level 3).

If we reject the existence of the Virtual and Quantum levels, the electron−positron annihilation and photon pair creation must be interpreted as the electron and positron being annihilated into the vacuum, with two photons emerging ex nihilo. Such an interpretation is physically and philosophically inconsistent, highlighting the necessity of recognizing intermediate ontological levels.

To clarify and provide coherence with the levels of reality, let us consider an example at the Quantum Vacuum, or Level 0 of reality.

### 4. Particles Creation in an Expanding Universe

The quantum vacuum is a state of a quantum field in which no detectable energy exists — that is, its energy level corresponds to the zero-point energy. However, this state is not perfectly static; it exhibits small fluctuations around the zero-point level, known as quantum vacuum fluctuations. From the particle perspective, the vacuum can also be understood as a state in which no observable particles are present. The quantum vacuum is an observer-dependent notion rather than an absolute state. A state that appears as empty in Minkowski space-time to an inertial observer may be perceived as populated by particles for an accelerated observer, manifesting as the Unruh effect. In this framework, the vacuum in Minkowski coordinates transforms into a thermal-like state in Rindler coordinates, illustrating that particle content is observer-dependent. Analogous transitions occur in other contexts: near black hole horizons in Hawking radiation, under strong electromagnetic fields in the Schwinger effect, and in expanding Friedman−Robertson−Walker universes. These phenomena reveal that the quantum vacuum does not represent an absolute zero-energy state. What is measured as zero energy by one observer may correspond to non-zero particle excitations for another. Bogoliubov transformations provide a formal framework demonstrating that the vacuum state of Minkowski spacetime, as seen by an inertial observer, manifests as a particle-populated state for an accelerated observer or within an expanding cosmological background. The excitation of the vacuum can be interpreted as an effective increase in the mass of off-shell virtual particles until their mass reaches the threshold satisfying Einstein's mass−energy relation. At this point, the vacuum state transitions into a particle-bearing state.

To investigate particle creation in an expanding spacetime, we consider the

Friedmann–Robertson–Walker (FRW) metric. The line element of this spacetime is written as:

$$ds^2 = dt^2 - a^2(t)dX^2$$

where a(t) is the scale factor and X represents the comoving spatial coordinates.

In flat (static) Minkowski spacetime, the Klein–Gordon equation for a scalar field φ is:

$$\left(\frac{\partial^2}{\partial t^2} - \nabla^2 + m^2\right)\varphi = 0$$

To generalize this to an expanding background, we introduce the conformal time parameter η defined by:

$$d\eta = \frac{dt}{a(t)}$$

With this change of variable, the FRW metric becomes:

$$ds^2 = a^2(\eta)(d\eta^2 - dX^2)$$

In this metric, the Klein–Gordon equation takes the form:

$$\ddot{\varphi} + 2\left(\frac{\acute{a}}{a}\right)\dot{\varphi} - \nabla^2\varphi + a^2 m^2 \varphi = 0$$

(where primes denote derivatives with respect to the conformal time η).

To simplify this equation, we define a new field variable:

$$\chi(\eta, X) = a(\eta)\varphi(\eta, X)$$

Substituting this transformation into the Klein–Gordon equation yields:

$$\ddot{\chi} - \nabla^2 \chi + \left[a^2(\eta)m^2 - \frac{\acute{a}}{a}\right]\chi = 0$$

This expression shows that the field χ behaves as a field with a time-dependent effective mass:

$$m_{eff}^2 = a^2(\eta)m^2 - \left(\frac{\acute{a}}{a}\right)$$

As the universe expands a(η) increases, and thus the effective mass m_eff evolves with time. When m_eff grows sufficiently so that the particle satisfies the Einstein relation $E^2 = m^2 + P^2$, a virtual (off-shell) particle can transition into a real (on-shell) particle. Therefore, the cosmic expansion can be interpreted as a mechanism for the transition from the Virtual Level (Level 1) to the Quantum Level (Level 2) of reality — that is, from a state of unobservable virtual fluctuations to one where real, detectable particles emerge. In an expanding spacetime, Level 1 virtual particles gain effective mass until they satisfy Einstein's mass–energy relation, transitioning into Level 2 on-shell quantum particles. This illustrates how spacetime expansion drives the continuous passage from potentiality to quantum actuality.

## 5. Nature of Vacuum; Hawking Radiation and the Schwinger Effect

We now focus on Level 0, the quantum vacuum, and explore examples of transitions from Level 0 to Level 1. The quantum vacuum should not be interpreted merely as a state of zero energy or an absence of particles. Rather, it is a composite entity composed of equal numbers of particles and antiparticles that cancel each other, producing the appearance of a zero-energy state from an external viewpoint. Analogous to the notion that zero can be expressed as the sum of +1 and −1, the quantum vacuum emerges from the symmetry between particles and antiparticles at Level 0. When this symmetry is disrupted and particle–antiparticle pairs separate, the process corresponds to the creation of virtual particles and the transition from Level 0 to Level 1. This process is mathematically represented in quantum field theory by the action of the creation operator on the vacuum state. Particles generated at Level 1 may later acquire sufficient conditions to satisfy the mass–energy relation and become on-shell, thereby transitioning to Level 2. Phenomena such as the Schwinger effect and particle production at black hole horizons (Hawking radiation) provide concrete realizations of these transitions.

In Hawking radiation and the Schwinger effect, particle–antiparticle pairs initially exist as virtual entities at the zero-energy vacuum (Level 0). In Hawking radiation, the intense gravitational field near a black hole separates the pair: the particle that escapes the horizon transitions to the virtual Level 1 and can subsequently acquire sufficient effective mass to reach Level 2, becoming observable, while the antiparticle remaining inside the horizon either annihilates or persists in the vacuum. In the Schwinger effect, a strong external electric field similarly separates the pair, promoting one particle to Level 1 and endowing it with the potential for subsequent actualization. In both phenomena, vacuum fluctuations under the influence of external fields or gravity give rise to real, observable particles. This process exemplifies a gradual ontological transition from Level 0 to Level 1 and ultimately to Level 2, illustrating that the quantum vacuum, despite its apparent zero energy, inherently possesses the capacity to generate actualized particles.

### 5.1. The Structure of the Quantum Vacuum and the Aharonov–Boh Effect

The structure of the quantum vacuum, or Level 0, is not limited to a simple symmetric arrangement of particles and antiparticles. From a structural perspective, the quantum vacuum exhibits a rich and highly nontrivial topology. One of the clearest manifestations of this complex structure is the Aharonov–Bohm effect, which reveals that a charged particle can be influenced by a magnetic vector potential even in regions where the magnetic field is zero.

The experimental setup is based on a modified double-slit experiment, as illustrated in Figure 3.

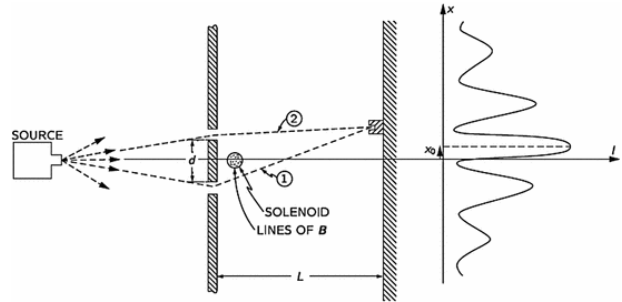

Figure 3; The Aharonov–Bohm effect(copied from the internet)

A solenoid carrying a confined magnetic field is placed behind the double-slit screen, so that the magnetic field outside the solenoid is zero, while the vector potential is nonzero. Classical physics predicts no effect on the electron interference pattern outside the solenoid, because the Lorentz force depends on the magnetic field , not the vector potential . Surprisingly, experiments show that the electron trajectories and interference pattern are shifted, demonstrating that has a real, observable effect. This phenomenon arises from the special topological properties of the configuration space. The solenoid effectively creates a "hole" in space, making it impossible to continuously contract all closed paths to a single point. Mathematically, these paths are classified into distinct equivalence classes labeled by integers. The Aharonov–Bohm effect thus provides clear evidence that the quantum vacuum is far from empty; spacetime itself possesses a nontrivial topological structure that can influence quantum particles even in regions devoid of local magnetic fields.

### 6. Revisiting Foundational Questions Through the Framework of the Four Levels of Reality

Having established the general framework of the four levels of reality within quantum field theory, we now turn to a re-examination of the fundamental questions that lie at the heart of both quantum mechanics and quantum field theory. By employing the hierarchical structure of the four levels, we aim to offer a renewed and coherent interpretation of these foundational issues, grounded in a deeper understanding of the ontological and physical relationships between the levels. In his article, Michael Redhead (1987) poses eight fundamental questions regarding Quantum Field Theory. Questions one through four fall within the scope of this paper. Questions three and four — concerning the nature of the vacuum and the essence of virtual particles, respectively — are addressed throughout the paper. However, we will attempt to answer the first and second questions, which are of more fundamental importance, in order.we aim to address these questions by leveraging the four-level ontological framework of reality.

### Q.1. Can QFT be given a particle interpretation and Indeed is there a formal underdeterminatin between field and particle approaches to the so called elementary particles?

Redhead frames the central question as whether the fundamental description of the physical world should be based on particles (individuals) or on fields (properties of spacetime points). In classical physics, a particle attains individuality either through its definite trajectory or an intrinsic identity that transcends its physical properties. In quantum mechanics, trajectories are undefined, so individuation via spatiotemporal continuity is impossible, leaving intrinsic identity (TI) as the only option. Experimentally, particle motion and the creation or annihilation of field excitations are indistinguishable. A common argument against a particle interpretation invokes quantum statistics: in quantum theory, exchanging two identical particles does not yield a physically distinct state, suggesting a lack of TI and identifying particles as field excitations. Redhead critiques this view as premature, showing that quantum state space can be decomposed into symmetric and antisymmetric sectors. With a symmetric Hamiltonian, a system remains confined to its initial sector. Thus, the statistical weighting in quantum theory reflects dynamical constraints, not the absence of TI.

This analysis underscores that rejecting the particle interpretation is not conclusive. To address the first question, we examine the particle interpretation of quantum fields through the four-level reality framework. As Redhead emphasizes, the main challenge is attributing trans-empirical individuation (TI) to virtual and quantum particles at Level 1 and 2 reality. At Level 2, corresponding to quantum particles, definite spatiotemporal coordinates or trajectories cannot be assigned, unlike in classical mechanics where such assignments underpin particle individuation. Classical particle individuation relies on trajectory and spatiotemporal continuity. At the quantum level, however, individuation is realized differently: Level 2 quantum particles manifest effective individuation only at the point of measurement. Unlike classical particles, whose identity is continuously maintained through locality, quantum particles achieve localized individuation upon detection, legitimizing a particle interpretation in the quantum regime. Virtual particles, which remain undetectable under ordinary circumstances, possess intrinsic individuation that becomes manifest when they transition to Level 1 reality ("on-shell"), thereby acquiring the potential for spatiotemporal localization at the point of detection. In other words, their fundamental individuation ensures that they can attain definite positions when conditions permit actualization.

A historical difficulty arises with the Klein–Gordon equation, which is second-order in time and permits negative probability densities, seemingly complicating particle interpretation. In contrast, the Schrödinger equation is first-order in time, ensuring definite positive densities and a straightforward particle view. This is resolved in QFT by recognizing that the Klein–Gordon field includes both positive and negative energy solutions, corresponding to creation and annihilation operators. The Schrödinger framework effectively omits annihilation, making particle existence implicit. Crucially, during interactions, whether involving virtual or real created particles, the eigenvalue modulus of the creation operator exceeds that of annihilation, ensuring positive Klein–Gordon probability densities . Hence, particle interpretation remains consistent. This analysis demonstrates that within the four-level framework, both virtual and real particles can be coherently interpreted as individuals, reconciling classical intuitions of particle identity with the formal structure of quantum field theory.

### Q.2. Does QFT resolve the problem of wave-particle duality in quantum mechanics(QM)?

To examine the second question, we begin our analysis with the well-known double-slit experiment, illustrated in Figure 2. In this setup, a light source is positioned at point 1, facing a screen with two extremely narrow slits, labeled 2A and 2B. Behind this screen, a second screen is placed, which we denote as Plane 3.

To examine the second foundational question, we begin with the paradigmatic double-slit experiment, illustrated in Figure 2. A light source at point 1 illuminates a screen containing two narrowly spaced slits, labeled 2A and 2B, with a second screen positioned behind at level 3 to capture the resulting pattern. If light were purely particulate, no interference pattern would emerge on the detection screen. Conversely, if light were purely wave-like, an interference pattern would form. Empirical observations show the formation of interference fringes on the screen, leading early physicists to conclude that light exhibits wave-like behavior. Yet, in the photoelectric effect, light manifests particle-like properties, demonstrating discrete energy transfer events. The apparent contradiction between these two experimental outcomes prompted profound conceptual challenges regarding the true nature of light and matter.

Niels Bohr proposed a resolution known as the Principle of Complementarity. Introduced in the 1920s as a cornerstone of the Copenhagen interpretation, this principle asserts that quantum phenomena cannot be exhaustively described using a single classical analogy, such as "particle" or "wave." Both descriptions are indispensable but mutually exclusive: which aspect manifests depends entirely on the experimental context. Formally, if an experiment is arranged to reveal wave-like properties—such as interference—the particle-like information is lost. Conversely, attempting to determine precise particle trajectories erases the wave characteristics. Though seemingly contradictory, these dual descriptions are complementary, and only by considering both can one attain a coherent and complete representation of quantum reality. Philosophically, Bohr emphasized that quantum "physical reality" is intrinsically tied to the conditions of measurement, rather than existing independently of observation. Complementarity thus serves not merely as a theoretical constraint within quantum mechanics but as an epistemological principle: it defines the interplay between what we can know and the phenomena themselves.

The wave–particle issue in the framework of quantum field theory can be analyzed in terms of the influence of the possible states of a phenomenon on its realized, actualized state. In other words, if a phenomenon possesses multiple potential states, and only one of them is actualized, all possible states nonetheless exert an effect on the realized outcome. This influence is a purely quantum effect and pertains to the properties of Level 1 and Level 2 realities. As we shall see, the wave–particle problem, contrary to Bohr's complementarity principle—which interprets it as an epistemological issue—is in fact an ontological problem, directly tied to the hierarchical structure of the levels of reality.

Re-examining the double-slit experiment (Figure 4) from the perspective of quantum field theory and the four-level framework of reality, consider a photon at Level 2, emitted from the source at point 1.

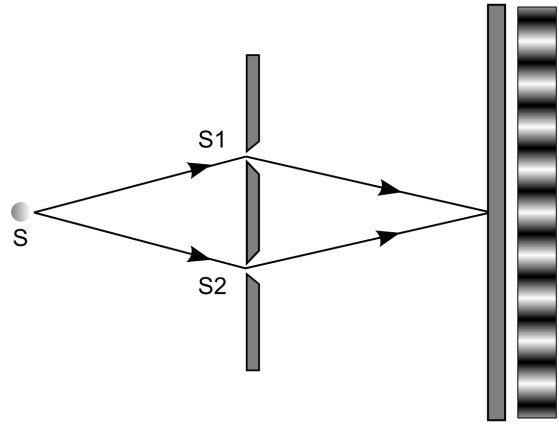

Figure 4; 2- slit experiment (copied from the internet)

This photon possesses two potential paths to reach the detector at point 3:

1- Source S → slit S1 → detector D

2- Source S → slit S2 → detector D

To calculate the probability of the photon arriving at detector D, one must consider the combined influence of both potential paths. Crucially, this represents an objective probability, grounded in the intrinsic structure of reality at Levels 1 and 2, and is fundamentally different from a subjective probability, such as a coin toss, which merely reflects an observer's epistemic limitations.

In quantum field theory, this objective probability is given by the squared modulus of the sum of the amplitudes associated with the possible paths. Were we to treat it as a subjective probability, as the sum of squared moduli of the separate amplitudes, the interference pattern would vanish. At the phenomenal level (Level 3), no wave-like interference would emerge, and light would manifest solely as a particle.

The photoelectric effect illustrates the complementary case: the experimental conditions ensure that the multiple possible paths interfere destructively, giving rise to a clearly particle-like manifestation of light.

From an ontological perspective, the double-slit experiment highlights that the potential states of a phenomenon actively shape the realized outcome. Unlike the epistemic interpretation suggested by Bohr's complementarity principle, the wave-particle duality is not merely about limitations of knowledge—it is a deep ontological feature of reality across levels. Objective quantum effects originate from the interplay of potentialities at Levels 1 and 2, which then manifest at Level 3 in accordance with the underlying structure of reality.

The apparent conflict of light's wave-particle duality is resolved within the framework of the four levels of reality. At the quantum levels (Levels 1 and 2), light inherently exhibits both wave-like and particle-like characteristics, and the realized outcome is objectively and ontologically influenced by all possible states. This explains the formation of interference patterns or particle-like behavior without reducing the phenomenon to the observer's subjective limitations. Bohr's principle of complementarity remains epistemologically valid, describing the observational constraints at the phenomenal level (Level 3), but the four-level analysis extends this understanding to an ontological dimension: the wave or particle nature of light exists independently of observation. By accounting for the influence of all possible states across these levels, the wave-particle tension is coherently resolved, providing an integrated picture of quantum reality.

Building on the analysis of the double-slit experiment and the role of possible quantum states, we can now summarize how the particle-wave duality can be consistently understood through the framework of the levels of reality. Quantum entities are fundamentally particles, but they exist as particles across different levels of reality, each with distinct properties. At Levels 1 and 2, the influence of all possible states on the actualized state ensures that the probability of occurrence is given by the vector sum of the amplitudes of all potential states. This intrinsic effect of potential states gives rise to wave-like behavior at the phenomenal level, Level 3. Conversely, if the experimental conditions are such that the possible paths cancel each other out, particle-like properties emerge in the observable phenomena. In this way, the apparent duality of quantum objects—wave or particle—is naturally reconciled without invoking a purely epistemic interpretation, providing a coherent ontological understanding grounded in the hierarchical structure of reality.

The four-level ontological model thus resolves the wave–particle duality and clarifies the objective nature of quantum probabilities as reflections of ontological potentialities rather than epistemic ignorance. Yet, this framework also echoes deep metaphysical intuitions that have appeared throughout the history of philosophy—from Aristotle's distinction between potentiality and actuality to Avicenna's gradation of possible worlds, Leibniz's monads, and Kripke's modal semantics. To fully appreciate the philosophical implications of these parallels, we now turn to a historical and conceptual analysis of the idea of potential states and their actualization across these classical frameworks.

## 7. A Review of the philosophical History of the Debate

This section provides a brief overview of historical perspectives on potency, act, and possible states. The aim is not to directly compare the theory of levels of reality with these classical views, but to highlight the depth of human thought and the insight of great philosophical minds. Recognizing similarities between scientific theories and these ideas clarifies the path of inquiry while offering an experience of intellectual pleasure to share with readers.

Aristotle introduced potentiality and actuality to analyze motion, a central issue among ancient Greek philosophers. In Metaphysics, he defines potentiality as the capacity of something to become something else—for example, an acorn has the potential to become an oak tree. This idea can extend to entities at more fundamental levels of reality. Actuality is the realized state of potentiality, illustrated through matter and form: matter represents pure potential without actuality, and form represents realized actuality. Aristotle emphasizes hierarchies of potentiality and actuality, where progression from potentiality toward actuality drives emergence and development (Aristotle, Metaphysics).

Avicenna (Ibn Sina) introduced possibility and necessity to understand existence. A necessary being exists by its essence, while a contingent being relies on causes and may not exist. Possibility marks the potential for contingent beings to be realized, and necessity the ultimate actualization. The movement from potentiality to necessity, mediated by causes, structures the unfolding cosmos (Ibn Sina, Al-Shifa': Metaphysics)).

Suhravardi, the philosopher of illumination, framed potentiality and actuality within his theory of light and levels of existence. Potentiality is the latent capacity of beings, and actuality its realization. Possibility depends on light and higher-level influences. All beings—actual, possible, or necessary—are forms of light in hierarchical levels. The transition from possibility to actuality involves interactions across levels, leading to emergence and perfection (Suhravardi, Hikmat al-Ishraq).

Thomas Aquinas integrated Aristotelian philosophy with Christian theology. He saw potentiality as the inherent capacity of beings, and actuality as its fulfillment. Possibility and necessity reflect divine order: some realities are necessary from God's perspective, others contingent. Hierarchies of realization connect human potential to divine actualization (Aquinas, Summa Theologica, I, q.2, a.3).

Alfred North Whitehead viewed reality as events arising from potentialities. Each actualization realizes and transforms prior potentials, creating a continuum of becoming. Potentialities at one level become actualities at the next, reflecting emergence in modern physics. Even possibilities carry causal relevance in shaping future actualities (Whitehead, Process and Reality, 1929, p.18).

Gilles Deleuze reinterpreted potentiality through virtuality: real potentials that can be actualized in diverse ways, producing multiplicities. Actuality is one instantiation among many potentials. Virtual-actual dynamics create hierarchical realities, guiding the emergence of higher forms (Deleuze, Difference and Repetition, 1968, p.38). This aligns with contemporary ideas on complex systems and emergence, where potentiality actively shapes reality.

In sum, the reflections of Aristotle, Avicenna, Suhravardi, Aquinas, Whitehead, and Deleuze reveal a persistent philosophical concern with the movement from potentiality to actuality. Across different traditions and frameworks, this progression highlights the dynamic interplay between latent possibilities and realized realities. Recognizing these patterns enriches our understanding of both philosophical and scientific perspectives, emphasizing that the unfolding of potential into actuality remains a foundational principle in shaping the structure of the cosmos.

Conclusion:

The analysis presented demonstrates that quantum field theory, when examined through the framework of the four levels of reality, provides a coherent account of both particle individuation and the resolution of wave–particle duality in quantum mechanics. From the perspective of trans-empirical individuation (TI), quantum and virtual particles, though lacking definite spatiotemporal trajectories at lower levels of reality, manifest individuated identities upon measurement or interaction. This approach aligns with classical intuitions regarding particle identity while remaining fully consistent with the formal structure of quantum field theory.

Regarding the wave–particle duality, the four-level framework reveals that the wave-like and particle-like behaviors of quantum entities are not merely epistemic limitations of the observer but reflect ontological features intrinsic to quantum levels. All potential states of a phenomenon, regardless of whether they are actualized, exert objective influence on the realized state, giving rise to interference effects and wave-like manifestations at the phenomenal level. In this sense, Bohr's principle of complementarity, while epistemologically valid, is subsumed under a broader ontological understanding that integrates the intrinsic potentialities of quantum systems across levels of reality.

Finally, situating these insights within the historical discourse on potentiality and actuality—from Aristotle through Avicenna, Suhravardi, Aquinas, Whitehead, and Deleuze—underscores the enduring significance of the transition from possibility to realization in shaping the structure of the cosmos. The four-level ontological model not only provides a unified conceptual framework for interpreting quantum phenomena but also opens avenues for advancing theoretical developments in quantum physics and understanding the role of potentiality in the emergence of physical reality.